\documentclass[twocolumn]{aastex62}

\received{June 01, 2021}
\revised{July 13, 2021}
\accepted{July 22, 2021}
\submitjournal{ApJL}
\shorttitle{Ancient massive quiescent galaxy at $z\sim3$}
\shortauthors{Kalita et al.}

\begin{document}

\title{An ancient massive quiescent galaxy found in a gas-rich z $\sim 3$ group}

\correspondingauthor{Boris S. Kalita}
\email{boris.kalita@cea.fr}

\author[0000-0001-9215-7053]{Boris S. Kalita}
\affil{CEA, Irfu, DAp, AIM, Universit\`e Paris-Saclay, Universit\`e de Paris, CNRS, F-91191 Gif-sur-Yvette, France}

\author[0000-0002-3331-9590]{Emanuele Daddi}
\affiliation{CEA, Irfu, DAp, AIM, Universit\`e Paris-Saclay, Universit\`e de Paris, CNRS, F-91191 Gif-sur-Yvette, France}

\author[0000-0001-7344-3126]{Chiara D'Eugenio}
\affiliation{CEA, Irfu, DAp, AIM, Universit\`e Paris-Saclay, Universit\`e de Paris, CNRS, F-91191 Gif-sur-Yvette, France}

\author[0000-0001-6477-4011]{Francesco Valentino}
\affiliation{Cosmic Dawn Center (DAWN), Copenhagen, Denmark}
\affiliation{Niels Bohr Institute, University of Copenhagen, Jagtvej 128, DK-2200, Copenhagen, Denmark}

\author[0000-0003-0427-8387]{R. Michael Rich}
\affiliation{Department of Physics \& Astronomy, Univ. of California Los Angeles, PAB 430 Portola Plaza, Los Angeles, CA 90095-1547, USA}

\author[0000-0002-4085-9165]{Carlos G\'omez-Guijarro}
\affiliation{CEA, Irfu, DAp, AIM, Universit\`e Paris-Saclay, Universit\`e de Paris, CNRS, F-91191 Gif-sur-Yvette, France}

\author[0000-0002-4343-0479]{Rosemary T. Coogan}
\affiliation{Max-Planck-Institut für Extraterrestrische Physik (MPE), Giessenbachstr.1, 85748 Garching, Germany}

\author[0000-0001-8706-2252]{Ivan Delvecchio}
\affiliation{INAF—Osservatorio Astronomico di Brera, via Brera 28, I-20121, Milano, Italy}

\author[0000-0002-7631-647X]{David Elbaz}
\affiliation{CEA, Irfu, DAp, AIM, Universit\`e Paris-Saclay, Universit\`e de Paris, CNRS, F-91191 Gif-sur-Yvette, France}

\author[0000-0002-0466-1119]{James D. Neill}
\affiliation{1 California Institute of Technology, 1200 East California Boulevard, MC 278-17, Pasadena, CA 91125, USA}

\author[0000-0001-9369-1805]{Annagrazia Puglisi}
\affiliation{Centre for Extragalactic Astronomy, Department of Physics, Durham University, Durham, UK}

\author[0000-0001-7975-2894]{Veronica Strazzullo}
\affiliation{University of Trieste, Piazzale Europa, 1, 34127 Trieste TS, Italy}
\affiliation{INAF—Osservatorio Astronomico di Brera, via Brera 28, I-20121, Milano, Italy}
\affiliation{INAF - Osservatorio Astronomico di Trieste, via Tiepolo 11, I-34131, Trieste, Italy}



\begin{abstract}
\noindent
Deep ALMA and HST observations reveal the presence of a quenched massive galaxy within the $z=2.91$ galaxy group RO-1001. With a mass-weighted stellar age of $1.6 \pm 0.4 \,$Gyr this galaxy is one of the oldest known at $z\sim3$, implying that most of its $10^{11}\rm \, M_{\odot}$ of stars were rapidly formed at $z>6$--8. 
This is a unique example of the predominantly passive evolution of a galaxy over at least $3<z<6$ following its high-redshift quenching and a smoking-gun event pointing to the early imprint of an age-environment relation. At the same time, being in a dense group environment with extensive cold-gas reservoirs as betrayed by a giant Ly$\alpha$ halo, the existence of this galaxy demonstrates that gas accretion shutdown is not necessary for quenching and its maintenance.


\end{abstract}



\section{Introduction} \label{sec:intro}
Hydro-dynamical simulations and semi-analytical models have found it difficult to reproduce massive ($\gtrsim 10^{11}\,\rm M_{*}$) quiescent galaxies (QGs) beyond $\rm z \gtrsim 3$ \citep{steinhardt16, schreiber18b, cecchi19}, while observations have been pushing the redshift boundary by detecting such systems at $\rm z \sim 3-4$ \citep{gobat12,glazebrook17, schreiber18a, deugenio20a, valentino20, forrest20a, forrest20b, saracco20}. 
Most of the systems so far observed  belong to a population of recently quenched `post-starburst' (PSB) galaxies with their last star-formation episode occurring over the last $<0.8\,$Gyr. However, we might still be missing a fraction of older ($\gtrsim 1\,$Gyr) QGs \citep[][]{deugenio20b, forrest20b}. Moreover, there has not been any insight on the influence of the environment over such galaxies, as most of those studied are field objects. 

Studying a younger sub-sample of a larger population of high-z QGs may have influenced our understanding of how quiescence occurs and persists. Currently, for star-formation to be suppressed in the $z>2$ epoch which usually features gas-rich star-forming galaxies, 
there are a variety of possible channels. These include merger driven starbursts \citep{puglisi21}, AGN feedback \citep{brennan18}, gas strangulation \citep{peng15}, halo quenching \citep{feldmann15}, and morphological quenching \citep{martig09}. Hence, further probing the high-z QG population, including those in dense environments, will provide opportunities to investigate the earliest mechanics of quenching.   

With this in mind, we report the detection of an extremely red and old quiescent galaxy  within a dense environment at $z\sim3$. Throughout, we adopt a concordance $\Lambda$CDM cosmology, characterized by  $\Omega_{m}=0.3$, $\Omega_{\Lambda}=0.7$, and $H_{0}=70$ km s$^{-1}\rm Mpc^{-1}$. We use a Chabrier initial mass function. 
Magnitudes and colors are on the AB scale.

\section{Galaxy-D and RO-1001} \label{sec:ro-1001}
\begin{figure*}[!ht]
    \centering
    \includegraphics[width=0.98\textwidth]{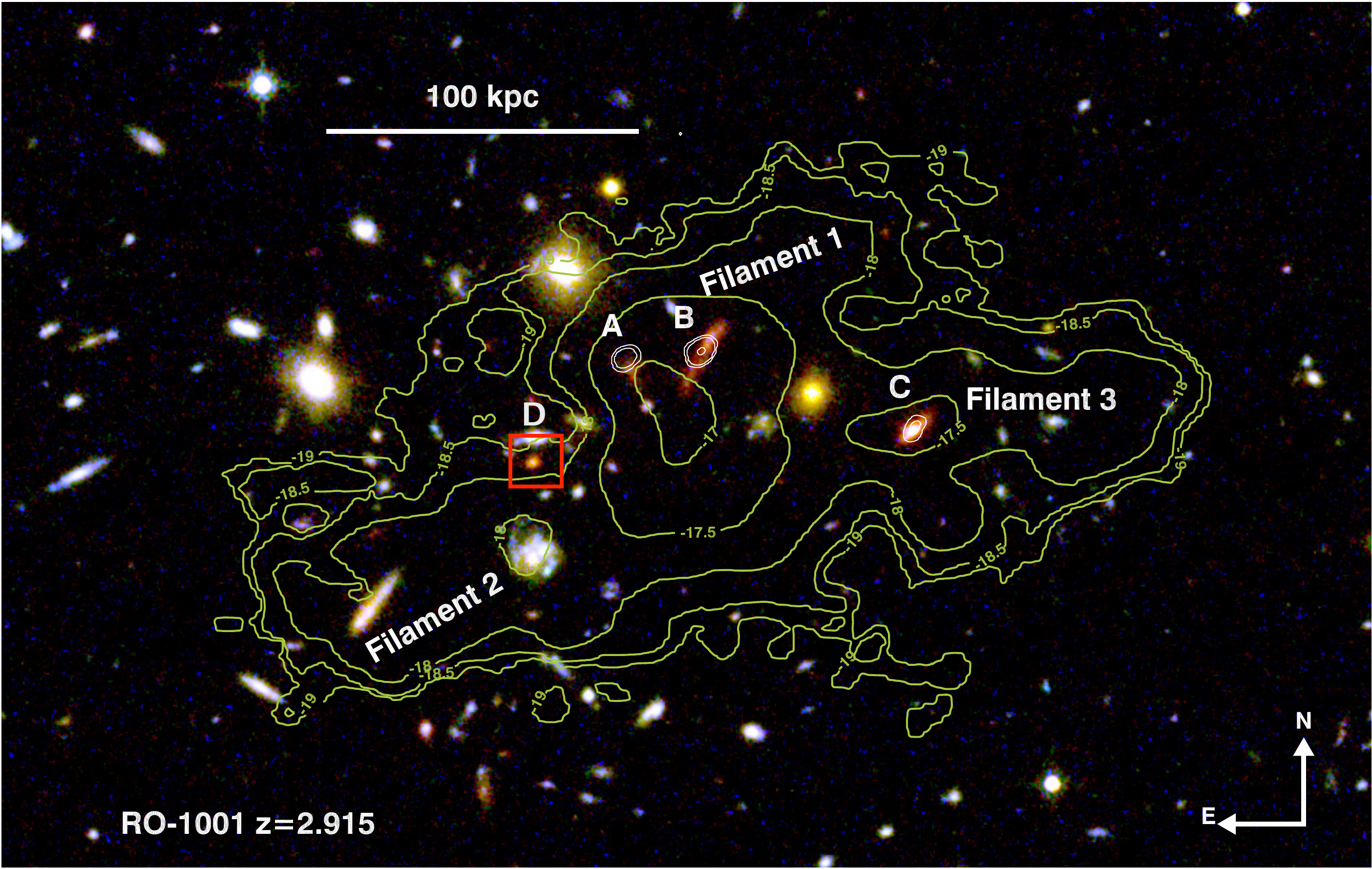}
    \caption{The core of RO-1001 as seen is HST/WFC3 (F606W, F125W, F160W). The green contours denote the Ly$\alpha$ nebula luminosity with the locations of the associated filaments possibly powered by cold-flow accretion  \citep[][]{daddi20}. Galaxy-D is marked with a red square.}
    \label{fig:ro-1001}
\end{figure*}

Three massive (log(M$_{\star}$/M$_{\odot}$)$\gtrsim$11) galaxies \citep[A, B, and C in Fig.~\ref{fig:ro-1001} and][]{daddi20} in the $z=2.91$ RO-1001 group have a combined star formation rate (SFR) $\sim 1250\,\rm M_{\odot}\,yr^{-1}$. NOEMA observations of the CO(3-2) transition showed that they lie at the same redshift of $z=2.91$. Furthermore, the presence of extensive cold gas reservoirs is suggested by Ly$\alpha$ observations that revealed extended filaments converging onto a single massive halo with a combined luminosity of $1.3 \pm 0.2 \times 10^{44}$ erg s$^{-1}$, most likely still affected by cold-accretion persisting in this system with estimated dark matter halo mass of $4\times10^{13}M_\odot$. These conclusions are based on the observed line profile, the velocity field, and the energetics underlying the Lya emission \citep[][]{daddi20}. Moreover, also pointing to the availability of cold gas is the redshift of the group, which places it in the epoch where cold flow accretion is expected to dominate \citep[][]{valentino15, overzier16}.
Galaxy-D was reported earlier as a possible passive member of RO-1001 \citep[Fig.~\ref{fig:ro-1001};][]{daddi20}. 

\section{Observations} \label{sec:observations}

\subsection{Optical and Near-IR imaging}
RO-1001 was observed with \textit{HST/WFC3} imaging in 3 bands (F160W, F125W and F606W) over a total of 11 orbits during Cycle 27 (Proposal ID: 15190, PI: E. Daddi). The data reduction was executed using the pipeline \textit{grizli} \footnote{https://github.com/gbrammer/grizli}. The $5 \sigma$ point-source sensitivities reached are 26.25 (F160W), 26.47 (F125W) and 26.39 (F606W) with a pixel scale of $0.06^{\prime\prime}$ and a half-power beam-width of $0.24^{\prime\prime}$ for F160W. Public COSMOS F814W imaging  
was also incorporated into the analysis. Due to a lack of coverage in z and y-bands, we include Subaru Hyper Suprime-Cam images were from the ``COSMOS2015'' database \citep[][]{laigle16}. Furthermore, we used the Ks band image from data release 4 of Ultra-VISTA. Finally, IRAC $3.6\,\mu$m and $4.5\,\mu$m images were taken from the Spitzer Matching Survey of the Ultra-VISTA Deep Stripes (SMUVS), while those at $5.8\,\mu$m were taken from the COSMOS2015 database. 

\subsection{Sub-mm imaging} \label{sec:ALMA_obs}
To have an estimate of the obscured star formation down to very faint levels in RO-1001, and as a result on Galaxy-D, we also obtained  sub-mm data. Atacama Large Millimetre Array (ALMA) band 7 observations were taken in Cycle 7 (Project ID: 2019.1.00399.S, PI: R.M. Rich). The data reduction was mainly carried out using the Common Astronomy Software Application (CASA). 
The final mosaics have a maximum sensitivity of $\rm 28\,\mu Jy\,beam^{-1}$, with a synthesised beam size of $0.49^{\prime\prime} \times 0.46^{\prime\prime}$.

\section{Analysis and results} \label{sec:analysis}
\subsection{Photometry} \label{sec:photometry}
\begin{figure}
    \includegraphics[width=0.47\textwidth]{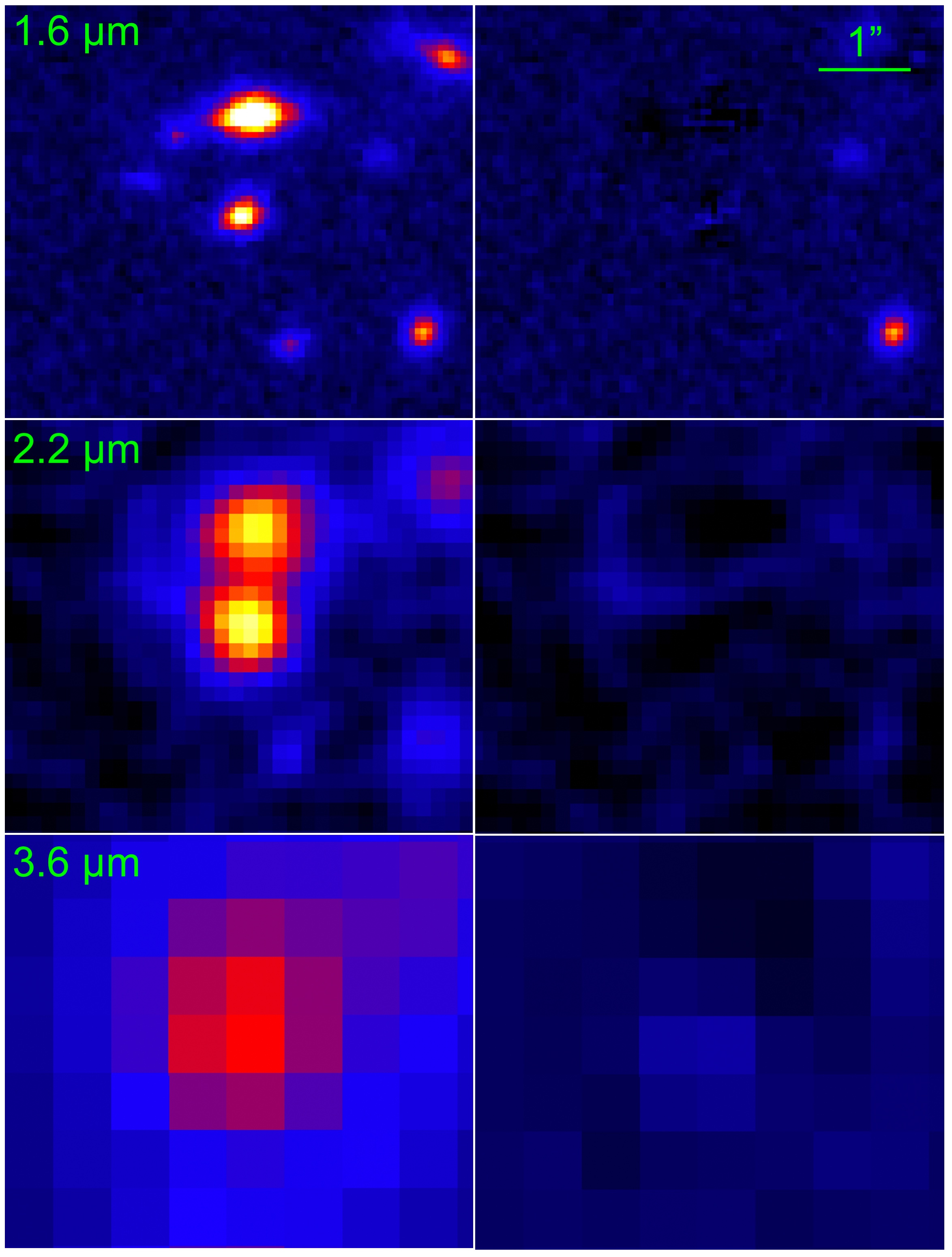}
    \caption{Panels showing Galaxy-D (at the centre) along with its surrounding foreground galaxies in 3 out of 5 wavelength bands for which GALFIT was used to extract the photometry. As detailed in Sec.~\ref{sec:photometry}, the results of H-band ($\rm 1.6\,\mu m$) were used to deblend at the other bands ($\rm 2.2\,\mu m, 3.6\,\mu m, 4.5\,\mu m, 5.8\,\mu m$, the first two of which are shown in the middle and lower rows respectively). More sources were fit and subtracted at longer wavelengths due to the increase in PSF size. The residual images are found to not have any fluctuations $>1 \sigma$.}
    \label{fig:galfit}
\end{figure}
In the COSMOS2015 catalogue \citep[][]{laigle16}  Galaxy-D is blended with a galaxy located 1$''$ to the North (Fig.~\ref{fig:ro-1001}). Hence, we make new flux measurements for it in each of the wavelength windows we have access to. We first use SExtractor for the 4 HST/WFC3 images since the the resolution is sufficient to clearly separate galaxy-D from its neighboring foreground galaxy. We use F160W as the detection image to carry out matched aperture photometry using isophotal radii. To ensure equivalent resolutions, we also convolve the F606W, F814W and F125W images with Gaussians such that their respective final point-spread-functions (PSF) have the same size as that in F160W. From the catalogue hence created, we find it to be undetected in F606W and F814W (which therefore provide upper-limits) while F125W and F160W have clear detections with signal-to-noise (S/N) of $\sim25$ and $65$ respectively. The z and y-bands also return non-detections, which are hence used to set flux upper-limits.

In order to ensure that all the flux from Galaxy-D is accounted for, we use galaxy morphological model fitting with the software GALFIT. Since the $1.6\,\mu$m (F160W) image has the highest S/N, we use it for this attempt. We fit s\'ersic profiles to both Galaxy-D and the foreground galaxy, along with Gaussian profiles to multiple surrounding objects (Fig.~\ref{fig:galfit}). We measure a total AB-magnitude of 24.00 for Galaxy-D and get a s\'ersic index of $2.0 \pm 0.2$. The error on the latter value includes the results from a co-addition of the F125W image to the F160W image to improve S/N as well as an independent analysis of the F125W image, all of which are in agreement. The s\'ersic index suggests that this galaxy might feature a disky component with a significant bulge component, within the scatter expected from high-z QGs \citep[][]{lustig21}. The galaxy is also found to be extremely compact with an effective radius, $\rm r_{e} = 0.14 \pm 0.03^{\prime\prime}$, which at z=2.9 is $1.1 \pm 0.1\,$kpc. We further use the flux difference between the GALFIT measurement and that from SExtractor, to scale up the catalogue (isophotal) flux measurement at $1.25\,\mu$m (F125W) as well as the upper-limits at $0.6\,\mu$m (F606W) and $0.8\,\mu$m (F814W). 

The morphological properties of Galaxy-D determined through the model fitting at $1.6\,\mu$m are further used as priors in Ks band ($2.1\,\mu$m) as well as the IRAC $3.6\,\mu$m, $4.5\,\mu$m and $5.8\,\mu$m. This is done in order to deblend the galaxy from its foreground neighbors, which are also simultaneously fit with models based on $1.6\,\mu$m priors (Fig.~\ref{fig:galfit}).  Images beyond this wavelength window do not show a detection, while the implied upper limits are not constraining, and are hence not used.   
    
\subsection{SED fitting} \label{sec:sed_fitting}
\begin{figure*}[!ht]
    \centering
    \includegraphics[width=0.49\textwidth]{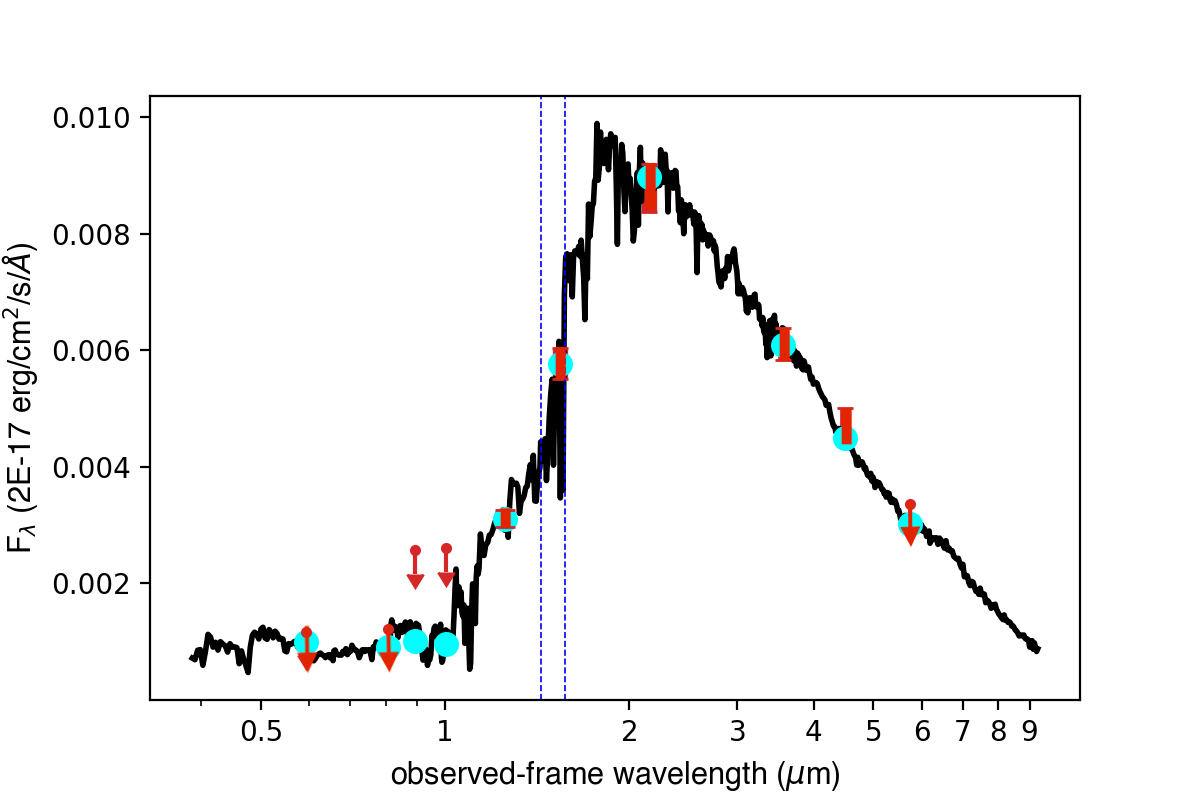}
    \includegraphics[width=0.49\textwidth]{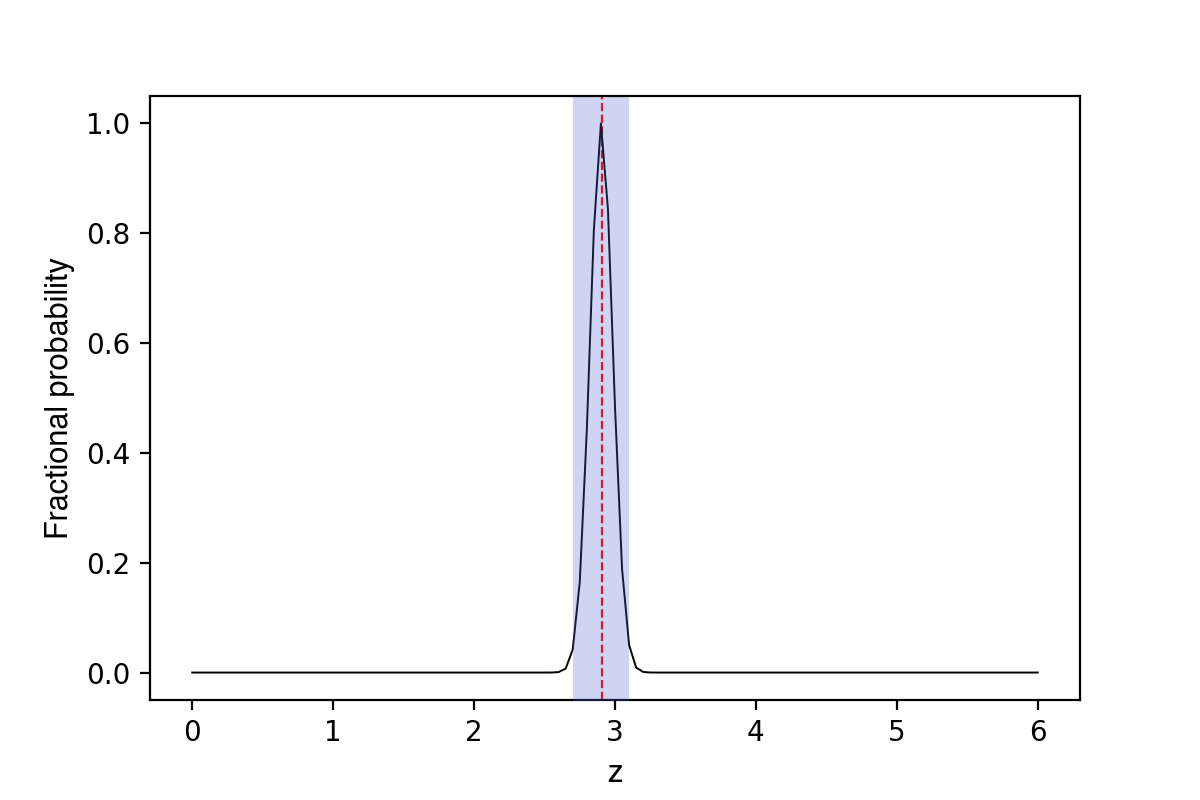}
    \includegraphics[width=0.70\textwidth]{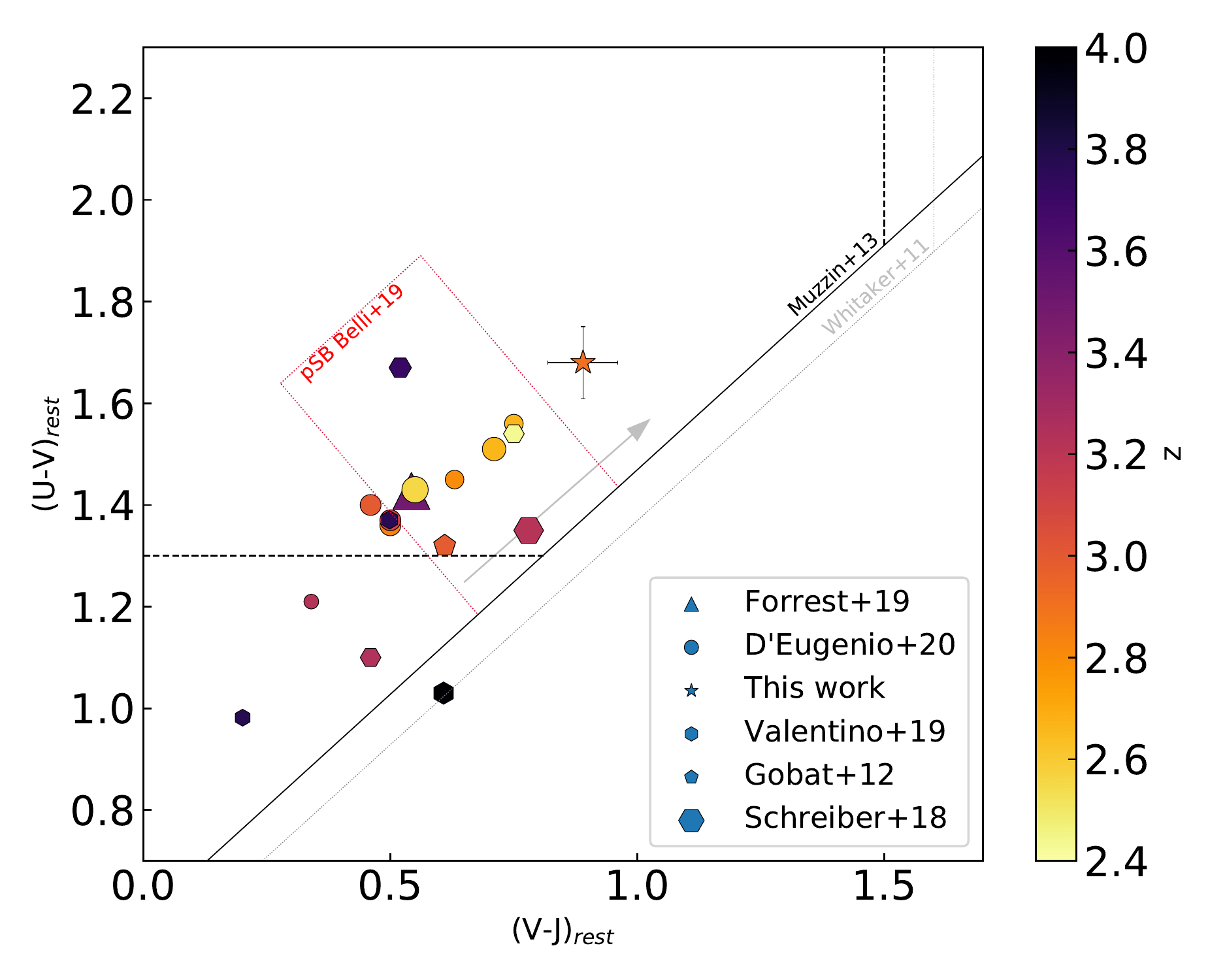}
    \caption{(Top-left) The best-fit SED template using the available photometry for Galaxy-D. The red bars show the observed photometry while the cyan points mark the expected values from the SED. The blue dashed-lines mark the locations of the Balmer and 4000\,\AA\ breaks. (Top-right) The probability distribution function for the photometric redshift of Galaxy-D. The red dashed line marks the spectroscopic redshift of RO-1001 while the blue shaded region marks the $90\%$ confidence region of the photometric redshift for Galaxy-D. (Bottom) The UVJ colors of high-z QGs ($\rm z>2.5$) adapted from \citet{deugenio20a}. The grey arrow shows the expected color evolution with increasing age. Galaxy-D is marked with a star, clearly outside the PSB region specified in \citet{belli19}. The marker sizes are proportional to the stellar mass.}
    \label{fig:sed_chi2}
\end{figure*}
We fit the photometry derived in the previous section with BC03 stellar population models \citep{bruzual03} to derive the properties of Galaxy-D (Fig.~\ref{fig:sed_chi2}, top-left). We use HYPERZ to determine the best fitting spectral energy distribution (SED) templates through a $\chi^{2}$ minimisation procedure. 

We investigate a redshift range of $0-6$ with a step size of 0.01.
For the BC03 templates, a range of metallicities are allowed ($\rm \frac{1}{5} Z_{\odot}$, $\rm \frac{1}{2}Z_{\odot}$, $\rm Z_{\odot}$ and $\rm \frac{5}{2}\,Z_{\odot}$). We also implement the \citet{calzetti00} along with SMC and LMC laws for dust attenuation, with extinction values, A$_{\rm V}=0-7$ in steps of 0.025. Finally, a wide variety of star formation histories (SFH) are used:   
\begin{enumerate}
    \item Simple stellar population (SSP): This is the most basic model with all of the stellar mass formed during an infinitesimal short burst of star formation at $t=0$. 
    \item Constant star formation models: Although we have enough evidence for the quiescence of the galaxy (discussed in detail later in Sec.~\ref{sec:quiescence}), we use this model featuring no decline in star formation (it remains constant throughout) to check if a high attenuation star-forming scenario is formally consistent with the available photometry. 
    \item Truncated models: These involve a constant SFR from $t=0$ up until a $t_{stop}$, beyond which $\rm SFR=0\,M_{\odot}\,yr^{-1}$. The $t_{stop}$ is varied from 0.1\,Gyr to a generous upper limit of $5.0\,\rm Gyr$ in steps of 0.1\,dex. 
    \item Delayed $\tau$-models: Finally, we use an SFH that has been widely preferred to model QGs. The exponentially declining SFH here is characterised as $\propto (t/\tau^{2})\,e^{-t/\tau}$ with a peak of star formation at $t=\tau$. The $\tau$ varies within [0.1, 5.0]\,Gyr with steps of 0.05\,dex.
\end{enumerate}
We investigate the reduced $\chi^{2}$ values from the fits using the various BC03 models, over the above mentioned grid of A$_{\rm V}$. We get the best fit with the delayed $\tau$-models (with a reduced $\chi^{2} \sim 0.4$), closely followed by SSP and truncated models. In comparison, we find a complete lack of fit using the constant star formation templates ($\Delta \chi^{2}>30$), with any level of reddening, reinforcing our claim of Galaxy-D being quiescent.

\subsection{Galaxy Parameters} \label{sec:galaxy_param}

The photometric redshift (photo-z) determined from the SED fitting is found to be $2.9 \pm 0.1$ (Fig.~\ref{fig:sed_chi2}, top-right), with the limits giving the $1\sigma$ confidence interval ($\Delta \chi^2<1$). We also derive a stellar mass, log($\rm M_{*}/M_{\odot})= 11.0 \pm 0.2$, in agreement with the previous results in \citet{daddi20}. Moreover, we find minimum levels of attenuation with the fitting procedure returning a value of $\rm A_{V} \sim 0.1$. We further confirm that our choice of gridding and `edge-effects' do not influence our results by investigating a variety of grids for $\rm A_{V}$ while ensuring that the parameter space on both sides of the best fit value is well sampled.

Following the analysis in similar studies \citep{gobat12,schreiber18b,valentino20,deugenio20a}, we define $t_{50}$, which is the time elapsed since the epoch of ``half-mass formation" up until the time of observation. We determine this quantity to be $1.6 \pm 0.4\, \rm Gyr$ at 90\% confidence, after marginalising on all the unknown parameters (SFH, metallicity, dust attenuation law and $A_V$, redshift). Fixing metallicity  to solar  does not change our results. Studying each of the values, we find that $t_{50}$ gets progressively lower as we go from sub-solar to super-solar metallicities. At $z=2.9$, for $\rm \frac{1}{5} Z_{\odot}$, we derive a best-fit $t_{50} \sim 2.2\,\rm Gyr$ which is almost equal to the age of the universe at that redshift, while for the super-solar $\rm \frac{5}{2} Z_{\odot}$, the $t_{50}$ is found to be $\sim 1.0\,\rm Gyr$. Although these results incorporate all previously mentioned SFH models together, using each model separately we find that the results from delayed-$\tau$ and truncated models are in agreement, while those using SSP (which can be considered as the light-weighted age) push the age upper-limit by $\sim 0.2\,$Gyr. Finally, fixing the redshift to that of RO-1001 ($\rm z=2.91$) does not change the constrain on age estimate. 

\begin{figure*}[!ht]
    \includegraphics[width=0.5\textwidth]{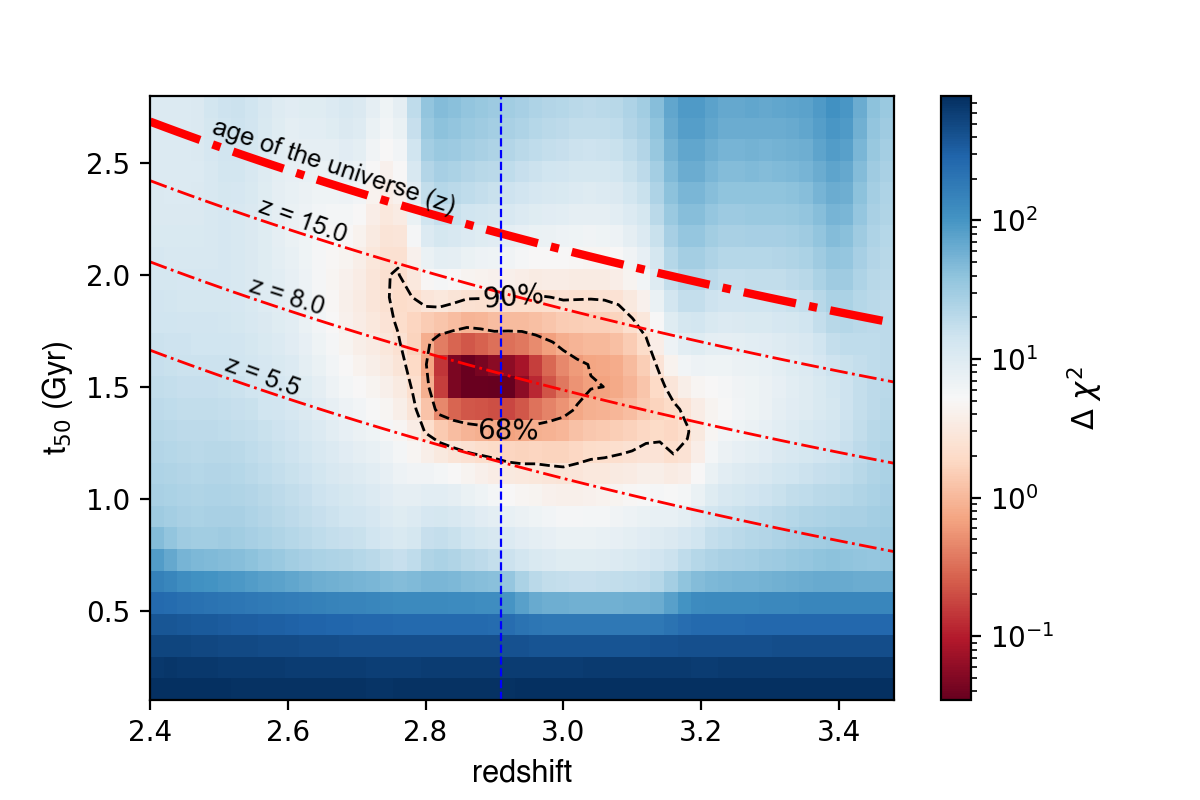}
    \includegraphics[width=0.49\textwidth]{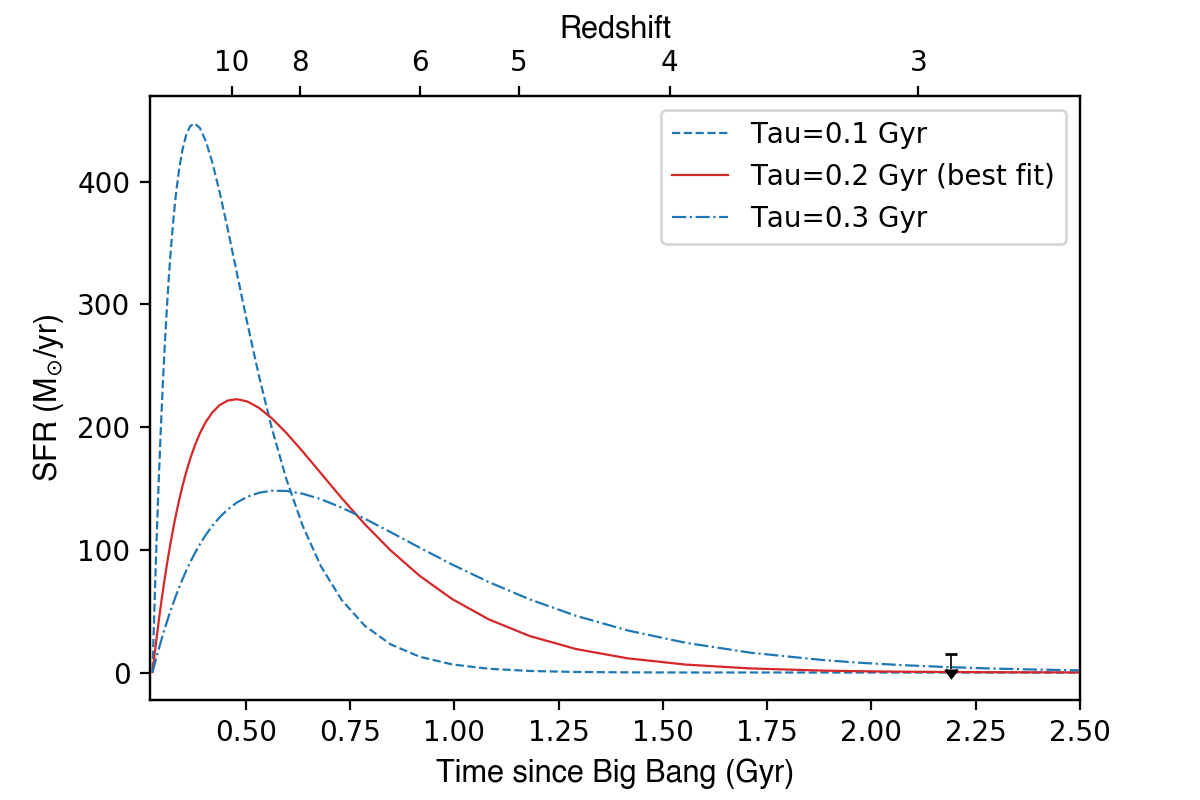}
    \caption{(Left) The variation of $\Delta \chi^{2}$ in the stellar population age ($\rm t_{50}$) vs redshift space. The dot-dash lines demarcate the lookback times for various redshifts along with the thickest of them showing the age of the Universe, all of which are functions of redshift (Right) The best fit delayed $\tau$-model SFH ($\tau=0.2\,$Gyr) as a function of age of the universe as well as redshift. Also shown are the SFH models with $\tau=0.1,0.3\,$Gyr which are within the 90\% confidence interval. Finally, the SFR $2\sigma$ upper-limit of $\rm 13\,M_{\odot}\,yr^{-1}$ derived from the ALMA $870\,\mu$m non-detection is shown with the black downward arrow.}
    \label{fig:age_chi_sfh}
\end{figure*}

We also investigate the robustness of our measurements with simulations. We perturb the flux measurements  with values drawn within Gaussians with sigma equal to the $1\sigma$ uncertainty of the respective flux measurement to generate 1000 sets of photometric datapoints. 
Besides a consistent average best-fit age of $1.7\,$Gyr, we find the $1\sigma$ scatter to be $0.15\,$Gyr (once accounting for the gridding effect). This uncertainty is slightly lower than the $0.2\,$Gyr ($1\sigma$ or $68\%$ confidence interval; Fig.~\ref{fig:age_chi_sfh}, left) that we get from the $\Delta \chi^{2}$ analysis. We repeated the exercise perturbing the best fit SED photometry instead of the observed photometry, finding the same results. Furthermore, for both cases, we also check the possible consequences of asymmetry in the range of acceptable ages in each fit ($\Delta \chi^{2} < 1$) by taking an average of their values. This also returns an age of $1.7\,$Gyr. Hence, we do not revise our more conservative results from the latter technique quoted throughout this work.

Furthermore, in order to have a SFH model-independent assessment of Galaxy-D within the framework of QGs at high-z, we also plot the (UVJ)$_{\rm rest}$ colors (Fig.~\ref{fig:sed_chi2}, bottom). These are determined using the F160W, Ks and IRAC $4.5\,\mu$m photometry, that match nearly perfectly to the rest-frame U, V and J bands, with small correction for the residual differences in effective wavelengths adopted from the best fitting SED. 

\begin{figure*}
    \includegraphics[width=0.94\textwidth]{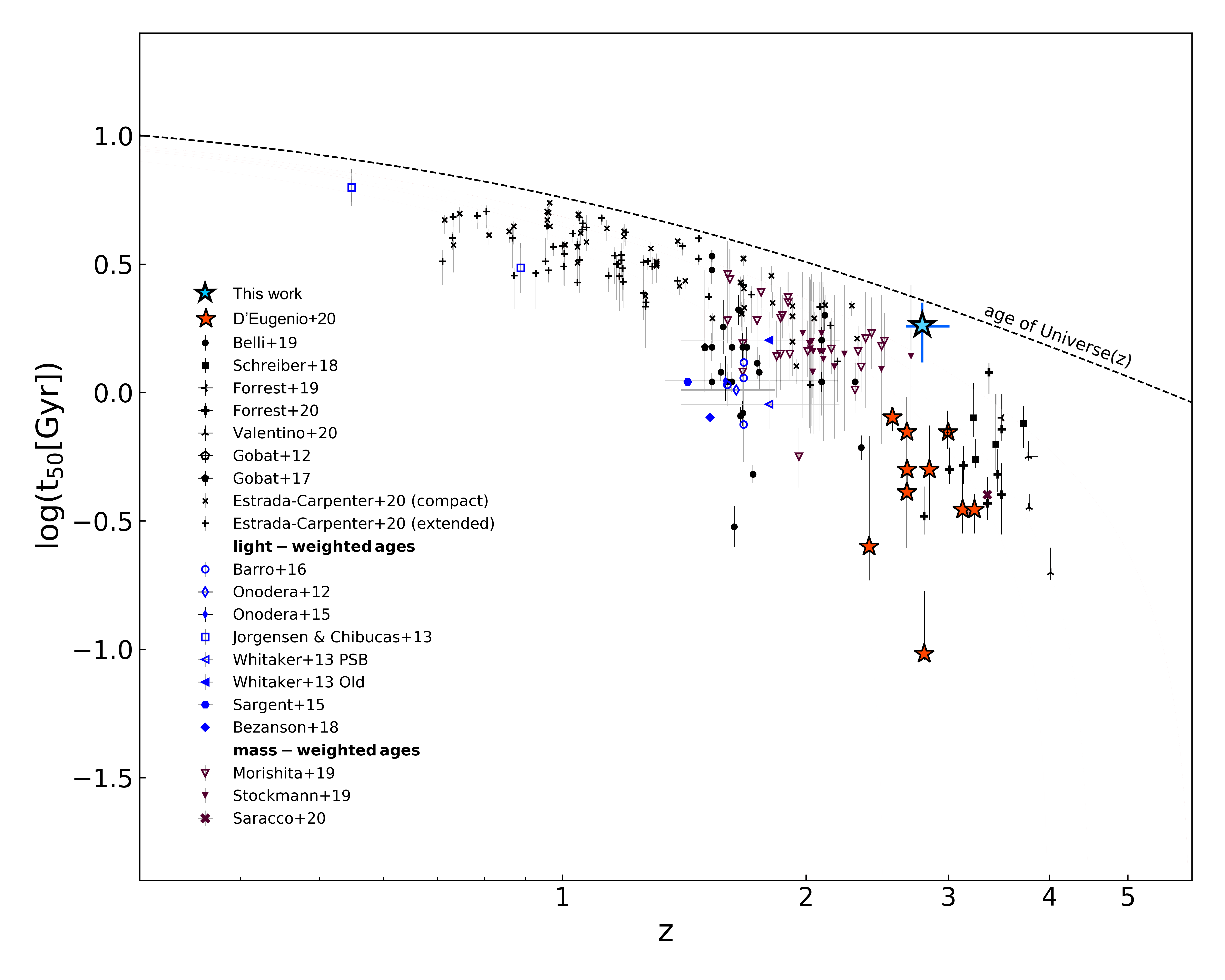}
    \caption{The stellar population age ($\rm t_{50}$) vs redshift distribution for log(M$_{*}/\rm M_{\odot}) > 10.5$ QGs adapted from \citet{deugenio20b}, with the references therein. Although the definition of the `mass-weighted ages' for a subset of the works add a level of ambiguity, they are found to be in agreement with our definition of $\rm t_{50}$ within a maximum of $0.1\,$Gyr (although in case of \citet{morishita19}, their `mass-weighted age' is equal to $\rm t_{50}$). We add Galaxy-D (light-blue star) with the $90\%$ confidence intervals shown as the errorbars for both axes.}
    \label{fig:age_z}
\end{figure*}

\subsection{Uniqueness and RO-1001 membership} \label{sec:group_membership}

The photo-z measurement of $2.9 \pm 0.1$ which is in remarkable agreement with the z$_{spec} = 2.91$ of RO-1001, strongly suggests group membership. This is further reinforced by the vanishingly low probability of $\sim 3 \times 10^{-5}$ for a chance alignment of Galaxy-D within $10^{\prime\prime}$ from the center of the RO-1001 group (as defined by the peak of the Ly$\alpha$ emission). This is estimated by the implied surface density of the 20 galaxies with H-band magnitudes $\leq 24.0$, J-H color $\geq1.1$ and H-Ks color $\geq2.2$ over the whole COSMOS area \citep{laigle16}. The color-selection thresholds determined by the photometry of Galaxy-D also incorporate the uncertainties in the measurements. However, 16/20 of these galaxies are consistent with being selected due to the $1\sigma$ scatter in their color measurements (recall that the COSMOS catalog contains more than half million sources). Hence, this number of objects found can be considered as a generous upper-limit. The lack of deep HST data like those we have for Galaxy-D primarily influences this conclusion. Out of the remaining 4/20, two lack photometry in most of the COSMOS observation bands and therefore do not have a photo-z reported in the catalogue. The final two galaxies (Laigle IDs: 749600 and  824982 \footnote{RA, dec: (149.47703, 2.43692) and (149.52635, 2.55155)}) are the only objects in the 2-deg-square area of COSMOS, which have been robustly found to have similar observed characteristics within the framework of the state-of-the-art multi-wavelength database available for the field. Also as a consequence, these galaxies have similar redshift, mass and age estimates as for Galaxy-D. However, due to a lack of sub-mm data (and therefore a confirmation of quiescence) along with any environmental information, a thorough study for these objects cannot be currently undertaken. However, this lack of counterparts should not be regarded as a suggestion of similar objects being absent. Rather, this could simply point to the inefficiency of current surveys to detect such galaxies with a high completeness, making Galaxy-D extremely rare within the current detection limits of surveys.   

Hence, besides being very unique, it is extremely unlikely that this galaxy is at a redshift different from that of the group RO-1001. Although, it cannot be currently determined with certainty whether Galaxy-D is exactly within the core of RO-1001 or part of the associated large-scale structure. In either case Galaxy-D can still be considered to be within a dense environment.

\section{Discussion} \label{sec:discussion}

\subsection{Quiescence and the last epoch of star-formation} \label{sec:quiescence}
Confirmation of quiescence at such high redshifts can be challenging \citep{glazebrook17, schreiber17, simpson17}. However for galaxy-D, the optical and NIR colors suggest a lack of  star-formation based on a complete disagreement with the star-forming SFH models (Sec.~\ref{sec:sed_fitting}).
Taking into account all possible SFH models (within 90\% confidence interval), we determine a SFR upper-limit of $4\,\rm M_{\odot}/yr$, $\times30$ below the MS.

The non-detection in our deep ALMA $870\,\mu$m data (Sec.~\ref{sec:ALMA_obs}) is used to determine a $2\sigma$ upper-limit SFR (conservative; fully accounting for its spatial extent of NIR emission of the galaxy based on its S\'ersic profile) of $13\,\rm M_{\odot}/yr$ assuming conservatively a main-sequence template appropriate for $z=3$ \citep{bethermin12}. This is already a factor $\sim 10$ below the main sequence at $\rm z \sim 3$ \citep{santini17}. The limit would be a further factor of 2 deeper if assuming instead a colder IR SED following \citet{gobat18}, i.e. at $\times20$ below the MS. Hence, it can be unambiguously concluded that Galaxy-D is truly quiescent, consistently from both the optical and IR side. 

Based on the mass-weighted ($t_{50}$) as well as light-weighted age limits (Sec.~\ref{sec:galaxy_param}), the primary episode of star-formation had already elapsed by $z\gtrsim6$ and most likely by $z \sim 8$ (Fig.~\ref{fig:age_chi_sfh}, left), right after which it likely experienced a rapid decline in its SFR characterised by the best-fit delayed $\tau$-model with $\tau=0.2\,$Gyr (Fig.~\ref{fig:age_chi_sfh}, right). However, this should be considered as an approximate upper limit, since $\tau=0.1\,$Gyr and the SSP (which has $\tau$ equivalent to 0) also return acceptable fits (within an overall $\Delta \chi ^{2} < 1$ or 68\% confidence interval). 

This form of rapid quenching has been extensively proposed to explain populations of QGs at $\rm z \sim 3-4$ \citep[e.g.,][]{schreiber18b, forrest20a, forrest20b, valentino20, deugenio20a, deugenio20b}. However, these are usually found to be PSBs, with $t_{50} < 0.8\,$Gyr (Fig.~\ref{fig:age_z}). Such studies usually leave out galaxies similar to Galaxy-D that are expected to have an older ($>1\,$Gyr) stellar population, due to preferential selection of brighter targets. \citet{forrest20b} shows that galaxies with such ages would be considerably less bright (by $\sim 1-2$ magnitudes) in near-IR, compared to PSBs. For example, in comparison to Galaxy-D, the \citet{valentino20} and \citet{deugenio20b} samples in Ks band are on average brighter by 1.7 and 2.4 magnitudes, respectively. This is also the case for a very recent detection of a $\sim 0.6\,\rm Gyr$ old massive QG detected in a protocluster environment at $\rm z=3.1$, with a Ks band magnitude 1.3 brighter \citep[][]{kubo21}. This difference is also found to manifest in the SFH-independent UVJ diagram Fig.~\ref{fig:sed_chi2} (bottom), which clearly shows a marked separation in color and hence an inferred age between the well studied PSB population and Galaxy-D.

Hence, spectroscopic analysis of such very-old galaxies is much more difficult, with requirements of extremely long integration times. Possibly as a result, similarly old galaxies at $z\sim3$ have not yet been spectroscopically studied yet (Fig.~\ref{fig:age_z}). We do acknowledge the presence of a population of galaxies at $z \sim 2-2.7$ in Fig.~\ref{fig:age_z} with seemingly similar ages from \citet{morishita19}. But  this comparison is affected by systematics, as their age estimates from bursty star-formation histories would be lowered by a factor of $\sim 1.5-2.0$ if smooth SFH models are employed as in our study \citep[see Fig.~12 in ][]{morishita19}, further emphasizing the unicity of Galaxy-D in the panorama of known old galaxies at $z\sim3$. 

Larger photometric studies, also with higher uncertainties regarding galaxies being truly quiescent, primarily feature young stellar populations with an extremely small fraction of their samples having ages $>1\,\rm Gyr$ \citep[e.g.,][]{straatman14, carnall20}. We do have examples of galaxies at lower redshifts $z\sim0.5$--2 with old enough stellar populations suggesting that they should be already $>1$~Gyr old at $z>3$ \citep[][and references therein]{tacchella21}.  If those ages inferences are correct (it is increasingly difficult at lower-z to accurately project star formation histories at the earliest times), it is still possible that the progenitors of the lower-z oldest galaxies were not yet hierarchically assembled into a single galaxy by $z\sim3$ \citep{mancini19}. It is hence unclear whether  old massive galaxies at $z\sim3$ are rare due to them experiencing re-accretion of gas and returning back to a star-forming population, to not having yet assembled, or they have simply not yet been detected at these redshifts.
Future time-intensive studies would hence be required to complete the mapping of high-z QGs in this regime. A crucial diagnostic for such attempts would be the Balmer and $4000\,$\AA\ breaks characterised by D$_{B}$ \citep{kriek06} and D$_{n}$4000 \citep{balogh99} respectively. The transition of the D$_{B}$/D$_{n}$4000 ratio towards values $< 1$ is a tell-tale sign of ages $\gtrsim 1.0\,$Gyr \citep{deugenio20b}. As an example, we calculate D$_{B}$/D$_{n}$4000 = 0.94 from the best-fit SED template (Fig.~\ref{fig:sed_chi2}, top-left), suggesting a small value for D$_{B}$. This is in stark contrast to the PSB population, which is usually characterised by much stronger Balmer breaks and hence higher D$_{B}$/D$_{n}$4000 ratios \citep[e.g.,][find an average value of 1.53 for their sample]{deugenio20b}. Future spectroscopy, that should be within reach of JWST, will be needed to validate these inferences.

\subsection{Tracing the evolution}

The short characteristic time-scales (Fig.~\ref{fig:age_chi_sfh}) and compact nature of  Galaxy-D ($\rm r_{e} = 1.1 \pm 0.1\,kpc$ at $z=2.9$) can be explained by scenarios like mergers \citep[e.g.,][]{puglisi21} that drive gas into a compact central region of the galaxy, leading to rapid consumption of gas through star formation. This quickly exhausts the available gas, thereby quenching the system and leaving behind a compact QG. Compact star-forming cores possibly related to mergers are already observed in the three primary star-forming galaxies in RO-1001 which are nearly identical in mass and size to galaxy-D \citep[][Kalita, B. S. et al. in prep.]{daddi20}, but forming over 1000\,$\rm M_\odot$~yr$^{-1}$ in stars complessively. 

Furthermore, with 50\% stellar mass ($\sim 5 \times 10^{10}\,\rm M_{\odot}$) being formed at $\rm z\gtrsim6$, we can determine whether its progenitor could have existed based on the expected galaxy mass-function, mainly sensitive to star-forming galaxies, at such redshifts. To make a comparison, we estimate a log number density of $-7.2 \pm 0.5\,$Mpc$^{-3}$ by assuming a single detection within $2.6 < \rm z < 3.2$ over the COSMOS $1.7\,\rm deg \times 1.7\,deg$ area. Then a number density of $>5 \times 10^{10}\,\rm M_{\odot}$ possible progenitors at $\rm z>5.5$ is derived from the results of \citet{grazian15}. We find the Galaxy-D-based density to be almost an order of magnitude lower than the latter value determined from the galaxy mass-function. Including the two candidate `ancient' QGs discussed in Sec.~\ref{sec:group_membership} brings this down to a factor $\sim 3$, with the possibility of further decrease if a few of rest are found to also have similar properties. Nevertheless, it cannot yet be ascertained whether more galaxies with formation epochs at $\rm z\gtrsim6$ within the COSMOS field do actually exist at similar redshifts. We can conclude however, that at least $10\%$ of the $z>6$ star-forming galaxies with stellar mass $>5 \times 10^{10}\,\rm M_{\odot}$ can be expected to have undergone quenching by $z\sim6$ to form old QGs like Galaxy-D by $\rm z\sim 3$. 

\subsection{Quiescence in a dense environment}
 
Given the estimated $t_{50}$, not only we need a quenching mechanism that could form it very early, but any galaxy rejuvenation has to be prevented thereafter. This primarily involves curbing accretion of gas or preventing the gas from forming stars. Galaxy-D belongs to  RO-1001 or associated sub-halos at $z=2.91$ (Sec.~\ref{sec:group_membership}). This halo is expected to  be fed by copious cold gas accretion \citep{daddi20} still at $z=2.91$, all the more in its past evolutionary and assembly phases and up to the $z>6$ quenching of Galaxy-D that  inescapably happened in a progenitor halo in which accretion must have been prominent. This  shows that one may have quenching and keep fully quenched galaxies (that  remain so for over 1\,Gyr at $z>3$--6) even with large amounts of diffuse cold gas and in presence of cold accretion. This is evidence that environmental quenching scenarios like cosmological starvation \citep[][]{feldmann15} and gas strangulation \citep[][]{peng15} do not play a necessary role in quenching and/or maintaining the quiescence of high-z galaxies.   

One of the alternative channels for inhibiting star formation would be AGN-driven feedback \citep[][]{sanders88, matteo05, hopkins06, mccarthy11}. Although Galaxy-D is not hosting a detectable AGN \citep[][]{daddi20}, multiple works have suggested that there are enhanced AGN fractions in high-z massive QGs \citep[e.g.,][]{olsen13, aird19}. Hence, it is highly likely that this galaxy would have at least experienced episodes of AGN activity in the past, inevitable given its large stellar mass. These AGN episodes might have or not curbed star formation as in QSO-mode quenching. However, AGN radio-mode activity was most likely crucial to allow it to evolve passively for $>1$\,Gyr, as supported by the high radio detection rate of distant quenched systems \citep{gobat18,deugenio20b}. Furthermore, other scenarios like morphological quenching \citep[stellar bulges preventing the collapse of gas for star formation;][]{martig09} could also be at play to prevent any significant late-time star formation. 

But a  question still remains: what sets Galaxy-D apart from the other similarly massive but intensively star-forming galaxies in RO-1001? It appears to be simply at a later evolutionary phase, while its vigorously star-forming counterparts have grown their large stellar masses later and are still undergoing rapid growth. They might have an eventual fate as that of Galaxy-D \citep[similar to the scenario outlined in][]{puglisi21}, and be hardly distinguishable in the stellar population properties once reaching $z\sim0$ more than 10\,Gyr later.

While Galaxy-D has a mass-weighted age ($t_{50}$) of $1.6 \pm 0.4\,$Gyr, most other quiescent galaxies at similar redshifts or above are younger PSBs and can predominantly be characterised as field galaxies (due to a lack of association with known overdensities). Therefore, the age difference between Galaxy-D and the currently known PSB population $\sim 1.0\,\rm Gyr$ is one of the first known evidences of an age-environment relation at high-z. Interestingly, similar difference of 1\,Gyr between clusters and field have also been found for local populations, where this relation has already been established \citep[e.g.,][]{bernardi06, gallazzi21}. But unlike at low-z, the age offset at $\rm z \gtrsim 3$ is a sizable fraction of the stellar ages themselves, thereby manifesting as much more pronounced observational differences as has already been discussed in Sec.~\ref{sec:quiescence}. 

In any case, Galaxy-D satisfying such a relation suggests a rapid hierarchical buildup of mass in dense environments, earlier than in the field. This in turn leads to the quenching scenarios discussed in the previous sections, mostly driven by mass and internal processes, while still agreeing with our conclusions of direct environment-dependent quenching scenarios not being  necessary. 


\acknowledgments
\emph{Acknowledgements:} We would like to express our gratitude to Raphael Gobat for his valuable inputs. We also thank Gabriel Brammer for assistance with the use of grizli for the HST data reduction. R.M.R. and J.D.N. acknowledge support from GO15910. V.S. acknowledges the support from the ERC-StG ClustersXCosmo grant agreement 716762. F.V. acknowledges support from the Carlsberg Foundation Research Grant CF18-0388 “Galaxies: Rise and Death” and from the Cosmic Dawn Center of Excellence funded by the Danish National Research Foundation under then Grant No. 140. A.P. gratefully acknowledges financial support from STFC through grants ST/T000244/1 and ST/P000541/1. B.S.K. would like to thank William G. Hartley (ETH Zurich, Switzerland), for providing crucial software-related files that were needed for this work.  
This paper makes use of the following ALMA data: 2019.1.00399. ALMA is a partnership of ESO (representing its member states), NSF (USA) and NINS (Japan), together with NRC (Canada), MOST and ASIAA (Taiwan), and KASI (Republic of Korea), in cooperation with the Republic of Chile. The Joint ALMA Observatory is operated by ESO, AUI/NRAO and NAOJ. 

%



\software{SExtractor \citep{bertin96}, GALFIT \citep{peng02,peng10}, grizli \citep[v0.4.0;][]{brammer18}, CASA \citep{mcmullin07}, HYPERZ \citep{bolzonella00}}

\end{document}